\documentclass[12pt,preprint]{aastex}

\shorttitle{Unbound Outflows in Accretion Simulations}
\shortauthors{De Villiers et al.}
\usepackage{graphicx}

\begin{document}

\title{Magnetically Driven Accretion in the Kerr Metric III: 
 Unbound Outflows}


\author{Jean-Pierre De Villiers\footnote{Now at Dept.
of Physics and Astronomy, University of Calgary, 2500 University Drive NW,
Calgary, Alberta, T2N 1N4}, John F. Hawley}
\affil{Astronomy Department\\
University of Virginia\\ 
P.O. Box 3818, University Station\\
Charlottesville, VA 22903-0818}

\and
\author{Julian H. Krolik, Shigenobu Hirose}
\affil{Physics and Astronomy Department\\
Johns Hopkins University\\
Baltimore, MD 21218}

\email{vjpde@ucalgary.ca; jh8h@virginia.edu; 
jhk@pha.jhu.edu; shirose@pha.jhu.edu}

\begin{abstract}

We have carried out fully relativistic numerical simulations of
accretion disks in the Kerr metric. In this paper we focus on the 
unbound outflows that emerge self-consistently from the accretion 
flow. These outflows are found in the axial funnel region and 
consist of two components: a hot, fast, tenuous outflow in the 
axial funnel proper, and a colder, slower, denser jet along the 
funnel wall.  Although a rotating black hole is not required to 
produce these unbound outflows, their strength is enhanced by black
hole spin. The funnel-wall jet is excluded from the axial funnel 
due to elevated angular momentum, and is also pressure-confined by 
a magnetized corona. The tenuous funnel outflow accounts for a 
significant fraction of the energy transported to large distances 
in the higher-spin simulations. We compare the outflows observed 
in our simulations with those seen in other simulations. 

\end{abstract}



\section{Introduction}

Over thirty years ago jets entered the lexicon of astrophysical
phenomena when it was realized that the double radio sources observed
in some galaxies are the product of tightly collimated
bidirectional outflows.  Since then there
has been substantial progress in both jet theory and jet observations.
We now know that in addition to active galaxies, jets are found in
mass-transfer binaries, galactic microquasars, protostellar systems,
and pulsars such as in the Crab nebula.  Collimated outflows are
clearly a natural occurrence, not solely the result of some rare and
unusual circumstance in the core of a few active galaxies.

Much of the theoretical work on jets has concentrated on broad
questions of collimation, and properties and stability of the
propagating jet itself.  A more difficult issue seems to be how the
jet is formed in the first place.  While a variety of radiative and
hydrodynamic processes have been examined as possible jet launch
mechanisms, the consensus that has emerged is that jets are
fundamentally a magnetohydrodynamic (MHD) phenomenon, most often
associated with an accretion disk (as argued, for example, by Livio
2002).  This picture is further supported by the
knowledge that accretion disks are themselves fundamentally MHD;
accretion is driven by MHD turbulence resulting from the
magnetorotational instability (MRI; Balbus \& Hawley 1998).

No other known mechanism works as well
and as naturally as large-scale magnetic fields to both accelerate and
collimate an outflow.  Stresses in large-scale poloidal fields that
thread through the disk at the appropriate angles can
accelerate plasma directly into a strong outflow.  This
magneto-centrifugal acceleration is sometimes referred to as the
Blandford-Payne mechanism, after the influential paper of Blandford \&
Payne (1982).  Strong magnetic pressure gradients resulting from
amplified toroidal fields have also been proposed as an acceleration
mechanism.  Regardless of the mechanism that drives the initial
acceleration, a large-scale poloidal field or a strong toroidal field
can provide effective collimation of the resulting outflow.

In these scenarios the energy for the jet comes from the disk and the
accretion therein.  The rotation of the central compact object, however,
is also a source of energy, even if the central object is a black hole.
The Blandford \& Znajek (1977) mechanism, for example, envisions field
lines connected directly to a rotating black hole; the rotation drives
outward-going Alfv\'en waves.  Punsly \& Coroniti (1990) suggested an
indirect magnetohydrodynamic mechanism for accomplishing much the same
end, by having the fields anchored in plasma orbiting near the rotating
black hole.  In either case, some of the black hole's spin energy is
extracted and transported outward as a Poynting flux.

Numerical simulations have long played an important role in examining and
validating many aspects of jet physics.  The work of Shibata \& Uchida
(1985) provides one of the earliest examples.  They simulated an accreting
disk of rotating gas, transfixed by vertical magnetic field.  In this
simulation radial field formed in the disk by infall is twisted into
strong toroidal field, which provides strong vertical magnetic pressure
forces to drive an outflow that is collimated along the vertical field.
Since this pioneering effort, many simulations have been performed,
featuring improved resolution, additional physical mechanisms, longer
time evolution, larger spatial domains, and even full three dimensions.
However, most of these jet simulations have, like Shibata \& Uchida,
featured a large-scale poloidal magnetic field as part of the initial
conditions.  In some cases a further simplification has been made
by treating the underlying accretion disk as a boundary condition.
In such studies the focus is on the ability of this large-scale field to
accelerate and collimate gas supplied by the disk boundary (e.g., Meier
{\it et al.} 1997; Romanova {\it et al.}  1997, Ouyed \& Pudritz 1997; Krasnopolsky,
Li, \& Blandford 1999).  Details differ from simulation to simulation,
in such things as the assumed magnetic field and the properties of the
disk boundary conditions, but, taken as a whole, the results provide
strong overall support for the magnetic launch and collimation scenario.

There remain, however, several significant questions that cannot be
answered by simulations of this sort.  First, do jets always
require a net large-scale poloidal field?  Under what circumstances do
such fields develop?  Can they be generated by a dynamo operating
within the disk, or must net field be brought in from large radius and
concentrated near the center?  These questions must be addressed
with fully global accretion simulations that do not assume the
presence of a large-scale field as an initial condition.

In contrast to simulations that have focused on jets, the simulations
developed for this series of papers has been directed toward the
dynamical properties of the black hole accretion disk itself.  Using a
general relativistic MHD (GRMHD) code  we have investigated accretion
disk structures that form from an initial condition consisting of an
isolated torus of gas fully enclosing a weak initial magnetic field.
An advantage of such an initial condition is that it is independent of
the boundary conditions.  Rather than assuming a pre-existing
large-scale field configuration, we can study the circumstances under
which such large-scale fields might develop naturally in the accretion
flow.

The first paper of this series, De Villiers, Hawley, \& Krolik (2003;
hereafter Paper I) presented an overview of a series of 3D general
relativistic MHD simulations of Keplerian accretion disks orbiting Kerr
black holes (the ``KD'' simulations).  The second paper of the series
(Hirose {\it et al.} 2004; hereafter Paper II) discussed the overall
magnetic field configurations in the accretion flow.  Among the results
noted in Paper I was the appearance of funnel-wall jets in all
simulations.  Such jets were also observed in a previous simulation
using a pseudo-Newtonian potential rather than full GR (Hawley \&
Balbus 2002; hereafter HB02).   Kato {\it et al.} (2004) have also
carried out similar pseudo-Newtonian simulations and analyzed in
greater detail the jets that were produced.  More recently, jets were
noted in 2D axisymmetric GR simulations by McKinney \& Gammie (2004).
What is noteworthy in all these simulations is the natural emergence of
jets from accretion disks that did not have large-scale fields included
as an initial condition.  These results have the potential to greatly
improve our understanding of the link between accretion and large scale
outflows, since these jets arise {\it self-consistently} from the
accretion process.  The success---or failure---of any self-consistent
simulation to produce jets will help to define the range of
circumstances that determine jet production in astrophysical systems.

In this paper, we will focus on the properties of the funnel-wall jets
seen in the KD simulations.   Relevant background material from Paper I
is briefly outlined in \S2. In \S3 we analyze the unbound outflows
seen in the KD simulations.  We include an overview of the jets, and
the details of separate regions within the jets.  In \S4 we briefly
review the jets seen in other accretion simulations.  
In \S5 we provide a summary of these results.

\section{Overview of Simulations}

We solve the equations of ideal MHD in the metric of a rotating
black hole.  The specific form of the equations we solve, and the
numerical algorithm incorporated into the GRMHD code are described in
detail in De Villiers \& Hawley (2003; hereafter DH03a).  For reference
we reiterate here the key terms and the definitions of the primary code
variables.

We work in the Kerr metric, expressed in
Boyer-Lindquist coordinates, $(t,r,\theta,\phi)$, for which
the line element has the form, ${ds}^2=g_{t t}\,{dt}^2+2\,g_{t
\phi}\,{dt}\,{d \phi}+g_{r r}\,{dr}^2 + g_{\theta \theta}\,{d \theta}^2
+g_{\phi \phi}\,{d \phi}^2$. We use the metric signature $(-,+,+,+)$.
The determinant of the 4-metric is $g$, and $\sqrt{-g} =
\alpha\,\sqrt{\gamma}$, where $\alpha$ is the lapse function,
$\alpha=1/\sqrt{-g^{tt}}$, and $\gamma$ is the determinant of the
spatial $3$-metric. We follow the usual convention of using Greek
characters to denote full space-time indices and Roman characters for
purely spatial indices.  We use geometrodynamic units where $G = c =
1$; time and distance are in units of the black hole mass, $M$.

The state of the relativistic test fluid at each point in the spacetime
is described by its density, $\rho$, specific internal energy,
$\epsilon$, 4-velocity, $U^\mu$, and isotropic pressure, $P$.  The
relativistic enthalpy is $h=1 + \epsilon + P/\rho$.  The pressure is
related to $\rho$ and $\epsilon$ through the equation of state of an
ideal gas, $P=\rho\,\epsilon\,(\Gamma-1)$, where $\Gamma$ is the
adiabatic exponent.  For these simulations we take $\Gamma=5/3$, unless
otherwise indicated. The magnetic field of the fluid is described by
two sets of variables, the constrained transport magnetic field,
$F_{jk}=[ijk]\,{\cal{B}}^i$, and magnetic field $4$-vector
$\sqrt{4\pi}\,b^\mu = {}^{*}F^{\mu\nu}U_\nu$.  The ideal MHD condition
requires $U^\nu F_{\mu\nu} = 0$.  The magnetic field $b^\mu$ is
included in the definition of the total four momentum, $S_\mu =
(\rho\,h\ + {\|b\|}^2)\,W\,U_\mu$, where $W$ is the Lorentz factor.  
We define auxiliary density and energy functions $D = \rho\,W$ and $E =
D\,\epsilon$, and transport velocity $V^i = U^i/U^t$. We also define
the specific angular momentum as $l=-U_\phi/U_t$ and the angular
velocity as $\Omega = U^\phi/U^t$.

In Paper I, we presented results of a series of high- and
low-resolution simulations, the KD (Keplerian Disk) set of disk
models.  These models have an initial condition consisting of an
isolated gas torus orbiting near the black hole, with a pressure
maximum at $r \approx 25M$, and a slightly sub-Keplerian initial
distribution of angular momentum throughout.  The initial magnetic
field consists of loops of weak poloidal field lying along isodensity
surfaces within the torus.  The emphasis in this paper will be on the
high-resolution models designated KD0, KDI, KDP, and KDE.  These differ
in the spin of the black hole around which they orbit, with $a/M = 0$,
0.5, 0.9 and 0.998 respectively.  These models used $192\times
192\times 64$ $(r,\theta,\phi)$ grid zones. The radial grid is set
using a hyperbolic cosine function to maximize the resolution near the
inner boundary.  The inner boundary is at $r_{in}= 2.05\,M$, $1.90\,M$,
$1.45\,M$, and $1.175\,M$ for models KD0, KDI, KDP, and KDE,
respectively, and the outer radial boundary is set to $r_{out}=120 M$
in all cases. The $\theta$-grid ranges over $0.045\, \pi \le \theta \le
0.955\, \pi$, with an exponential grid spacing function that
concentrates zones near the equator; reflecting boundary conditions are
enforced in the $\theta$-direction.  The $\phi$-grid spans the quarter
plane, $0 \le \phi \le \pi/2$.

Although we will focus our analysis on the KD models, we also will draw
upon the results of a number of other models for comparison purposes.
Most of these other models are run at lower resolution, using
$128\times 128\times 32$ grid zones.  The effects of numerical
resolution will be gauged in part by a direct comparison between KDP
and the lower-resolution model KDPlr (Paper I).  We will examine the
series of models that began with constant angular momentum tori around
holes with $a/M = -0.998$, 0.0, and 0.9, the ``SF'' models described in
De Villiers \& Hawley (2003; hereafter DH03b).  We will also consider
the presence (or absence) of jets in several pseudo-Newtonian
simulations. Finally we will present some results from two new
simulations, a high-resolution simulation with $\Gamma = 4/3$, and a
disk with an initial toroidal field.

\subsection{Simulation diagnostics}

Three-dimensional numerical simulations generate an enormous amount
of data, only a representative sample of which can be examined.
Our analysis is based on a specific set of volume- and shell-averaged
history data, taken every $M$ in time,
and complete data snapshots taken every $80M$ in time.
In addition, the complete density field is output every $2M$
for use in making highly time-resolved animations.
Although the details of the history calculations are given in Paper I,
we provide here a brief summary to clarify the calculations of mass, 
energy, and angular momentum transported by the unbound outflows. The flux 
${\cal F}$ of a given quantity through a shell at radius $r$ is 
computed using
\begin{equation}\label{avgdef}
\langle{\cal F}\rangle(r) = \int\int{{\cal F}\,\sqrt{-g}\, d \theta\,d \phi},
\end{equation}
where the bounds of integration range over the full $\theta$ and $\phi$ 
computational domains. 
We evaluate the rest mass flux $\langle\rho\,U^r\rangle$, 
energy flux $\langle{T^r}_t\rangle$ and the angular momentum flux
$\langle{T^r}_{\phi}\rangle$. Since we are interested in material that leaves
the computational domain through the outer boundary, we will refer to these
fluxes as ejection rates when they are evaluated at the outer boundary, 
$r_{out}$.  Gas that leaves the grid at the inner radial boundary is
considered to be accreted into the black hole; we refer to this as
the black hole accretion rate.
We account for changes in the total mass, energy, and angular 
momentum in terms of the corresponding net fluxes that leave the grid,
either into the hole or out of the grid, by integrating over time,
\begin{equation}\label{surfflux}
\left\{{\cal F}\right\}_{\rm{out}} = \int{ 
 dt\,\langle{\cal F}\rangle}(r_{\rm{out}}),
\end{equation}
for example.

The data on radial fluxes found in the time-history data is insufficient 
for the present study. For example, not all of the gas leaving the grid 
at the outer boundary is unbound; to distinguish between bound and unbound 
matter, we impose on the integrands the restriction that the specific total 
energy satisfy $-h\,U_t > 1$ when seeking to include only unbound material.  
This can be computed only by using the complete data dumps made every $80 M$ 
in time.

One of the important issues in jet physics is the nature of the
forces that account for jet acceleration.  
To compute forces from the simulation data, we begin
with the conservation law $\nabla_\mu\,T^{\mu\nu}=0$, where the
energy-momentum tensor consists of a fluid part,
$T^{\mu\nu}_{(FL)}=\rho\,h\,U^\mu\,U^\nu+P\,g^{\mu\nu}$, and an
electromagnetic part,
$T^{\mu\nu}_{(EM)}={(4\,\pi)}^{-1}\left({F^\mu}_\alpha\,F^{\nu \alpha}
- F_{\alpha \beta}\,F^{\alpha \beta}\,g^{\mu \nu}/4\right)$. As
described in DH03a, the momentum equation used in the numerical solver
is obtained by applying the projection tensor, $h_{\alpha \nu} =
g_{\alpha \nu}+U_\alpha\, U_\nu$, to the conservation law. Doing this,
and then using the equations for conservation of energy
($U_\nu\,\nabla_\mu T^{\mu\nu}=0$) and baryon number ($\nabla_\mu
\rho\,U^\mu=0$), it is straightforward to show that the fluid portion
of the momentum equation can be rewritten as
\begin{equation}
 h_{\alpha \nu}\,\nabla_\mu\,T^{\mu\nu}_{(FL)} = 
 \rho\,h\,U^\mu\,\nabla_\mu U_\alpha + 
  \left(\nabla_\alpha P + U_\alpha\,U^\nu\,\nabla_\mu P \right)
\end{equation}
One can express the divergence of the electromagnetic part of the
energy-momentum tensor in terms of the Lorentz force,
$\nabla_\mu\,T^{\mu\nu}_{(EM)}=-J_\mu\,F^{\mu \nu}$ (Misner, Thorne, \&
Wheeler, 1973; hereafter MTW), where $J^\mu$ is the current 4-vector,
$J^\mu = {(4\,\pi)}^{-1}\,\nabla_\nu\,F^{\nu \mu}$.
Applying the projection tensor to this result, along with the ideal MHD
condition, $U_\mu\,F^{\mu \nu} = 0$, and grouping with the previous
result, we obtain the force law, 
\begin{equation}
\rho\,h\,U^\mu\,\nabla_\mu U_\alpha =
  -\left(\nabla_\alpha P + U_\alpha\,U^\nu\,\nabla_\nu P \right) +
  J_\mu\,{F^{\mu}}_\alpha, 
\end{equation} 
where the left-hand side represents the product of inertia and
4-acceleration acting on the fluid, and the right-hand side represents
the sum of the pressure-gradient and Lorentz forces acting on the
fluid. As pointed out in MTW, this is a very useful form of the
momentum equation, since the right-hand side expresses departure from
geodesic motion (i.e., the right-hand side is identically zero for
free-falling dust). However, in the present context this formulation
has the disadvantage that the centrifugal and Coriolis force terms are
hidden in the left hand side. Similarly, the ``J dot F'' term combines
contributions from both magnetic pressure gradients and Maxwell stresses;
in order to distinguish between these two contributions, it is
possible to evaluate the components of $\nabla_\mu T^{\mu\nu}_{(EM)}$
using the secondary code variables for the magnetic field, $b^\mu$.

The contribution due to pressure gradients can be expressed as a 4-vector
\begin{equation}\label{pequation}
{\cal F}^\mu_{(P)} = -\left(g^{\mu \nu}\nabla_\nu P + 
 U^\mu\,U^\nu\,\nabla_\nu P \right) = -h^{\mu \nu}\,\nabla_\nu P,
\end{equation}
an expression which can be readily evaluated using code variables. We can
similarly write an expression for the Lorentz force 4-vector, 
\begin{equation}\label{lfequation}
{\cal F}^\mu_{(L)} = h^{\mu \nu}\,J^\alpha\,F_{\alpha \nu}.
\end{equation}
Recall that $F_{\alpha \nu}$ are the CT magnetic field and EMFs
[eqns.(14) and (35) of DH03a], which are the fundamental magnetic field 
variables evolved in the code.  Evaluation of $J^\alpha$ requires data for 
three adjacent time steps, since time derivatives must be evaluated.  Such
datasets were generated from the late-time data dumps by direct evolution 
using the GRMHD code. 

\section{Unbound outflows in KD models}

The unbound outflow is, at its most basic level, that portion of the 
flow that is \textit{outbound} and \textit{unbound}.  Specifically, 
it is where the radial momentum is positive, and the specific energy 
at infinity is greater than unity, i.e. $-hU_t > 1$. These criteria 
are met in the axial funnel and the funnel wall regions of the simulation 
volume, and the unbound outflow can be divided into two sub-regions: a 
hot, tenuous, fast component in the funnel, and a colder, denser, slower 
component along the funnel wall. These two regions have distinctive 
properties. We turn to the dense funnel-wall component first.

\begin{figure}[ht]
    \epsscale{1.0}
    \plotone{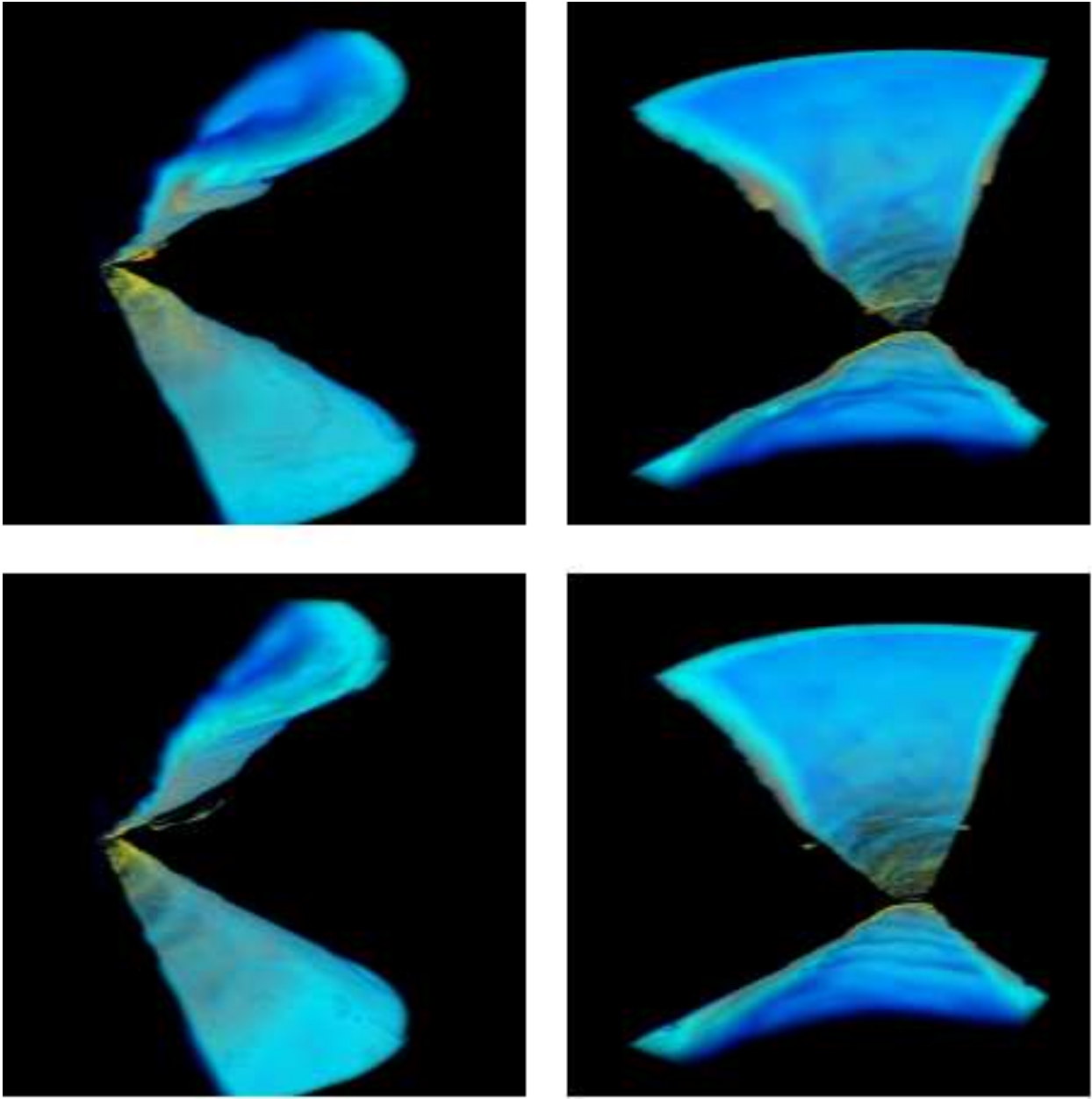}
    \caption{\label{Jetstills} 
    Still frames taken from an animation of the density $\rho$ in the
    jets for model KDP, i.e., where $U^r > 0$ and $-hU^t > 1$. The top
    row shows two views of the jet at $t = 7840\,M$; the bottom row is
    for $t = 8080\,M$.  One view looks outward from the polar axis, the
    other inward from outside the jet.  The shape corresponds to the
    quarter-plane $\phi$ coordinate range used in the simulation.  The
    outer boundary of the image is at $r \approx 60\,M$.  The density
    is plotted as a volumetric rendering using a logarithmic scale, where
    blues indicate densities in the range of ${10}^{-4}$ to
    ${10}^{-3}$, grey values near ${10}^{-2}$, and yellows and reds
    values in the range ${10}^{-1}$ to ${10}^{0}$ relative to a
    fiducial value taken at the base of the jet.}
\end{figure}

The basic structure of the funnel-wall component is 
illustrated by Figure \ref{Jetstills}, which consists of still 
frames taken from an animation of gas density $\rho_{jet}$, that is
the density $\rho$ where $-hU_t > 1$, for model KDP (which most
clearly exhibits the jet properties observed in the KD models). The 
animation frames show that the unbound density forms a biconical 
outflow lying along a narrow boundary between the largely evacuated 
axial funnel and the coronal envelope, hence the name ``funnel-wall 
jet'' introduced in Paper I. 

The funnel-wall jets are not transient features in these models.  
They are established early in the simulation at a time coincident with
the establishment of the quasi-steady accretion flow into the black
hole, and they endure for the length of the simulation.  We have
compiled extended animation sequences of the jets, and Figure
\ref{Jetstills} presents two views at two different times from such a
sequence, one view looking onto the jet from the funnel side of the
computational volume and the other from the corona side.  The quarter
hourglass shape corresponds to the coordinate range in $\phi$ used in
the simulation.  The complete animation sequence reveals that the
funnel-wall jet is highly dynamical, and its shape varies on a time
scale comparable to the orbital period of the main disk body.  This
variability is due in part to a changing dynamical balance between
forces in the magnetically-dominated funnel and the coronal envelope,
as well as variations in the rate at which matter is injected into the
jet.  The variable injection rate can be seen in the relatively dense
spiral features that appear near the base of the jet and propagate
outward, becoming more diffuse.  Density variations also appear as
episodic axisymmetric ``rolls'' that propagate outward; an example of
this is visible halfway up the upper jet in the top panels of the
figure.  Such rolls appear to be correlated with the arrival of
especially dense knots of material in the inner torus region; part of
this material is then expelled through the jet in a more or less
axisymmetric manner.

In contrast to the funnel-wall jet, the axial funnel component of the 
unbound outflow is largely invisible in density and mass flux plots. 
However, as noted in Paper I, the axial funnel contains a tenuous, hot, 
fast radial outflow with a significant electromagnetic component.

\begin{figure}[ht]
    \epsscale{1.0}
    \plotone{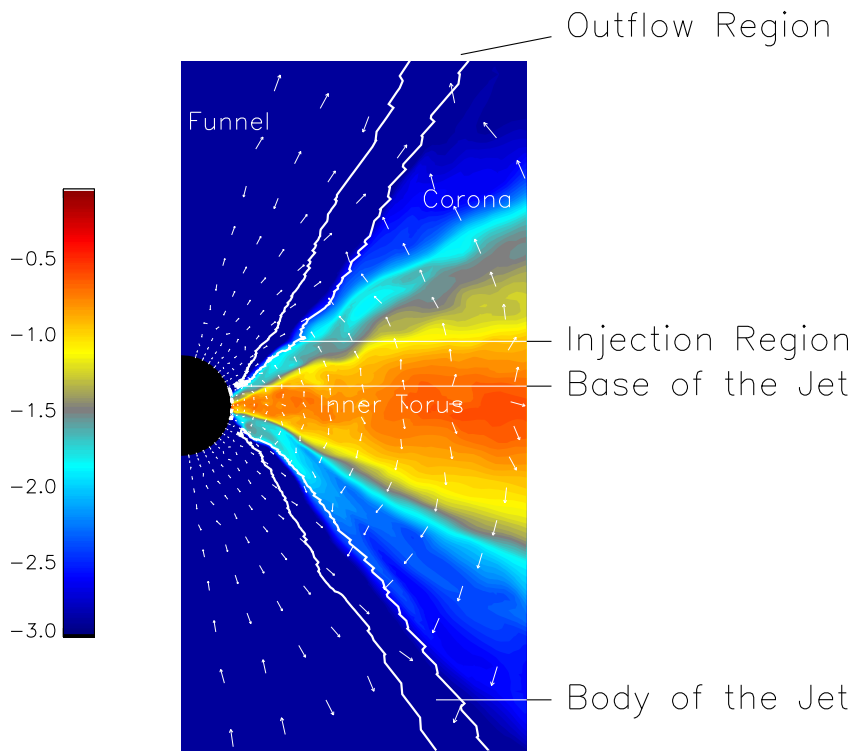} 
    \caption{\label{Overview} 
    General structural features of the jets. This plot shows a
    composite view of several azimuthally-averaged late-time quantities
    for model KDP.  In general, the jet is well-established after two
    orbital periods of the main disk, and the features shown here apply
    equally to all models that exhibit jets. The color contours show
    gas density ($\rho$) on a logarithmic scale. The boundary of the
    jet is indicated by a thick white contour (see text for
    explanation). The vector field indicates the direction, but not the
    magnitude of the net forces acting on the accretion system. The
    length of the arrows is scaled to the local zone width. Not all
    vectors are shown; a stride of 8 was used in both the radial and
    polar grids to reduce the density of the vector plot.
  } 
\end{figure}

To organize the discussion we divide the unbound outflow into distinct
regions, as suggested by the animations and from qualitative analysis.
For brevity, we will refer to the dense, cold funnel-wall jet simply as
``the jet'', and the hot outflow in the axial funnel as ``the funnel
outflow''. The various components of the unbound outflow are
illustrated in Figure \ref{Overview}, which shows an
azimuthally-averaged composite from model KDP.  The color contours
depict gas density on a logarithmic scale.  The boundary of the jet is
indicated by a thick white contour; this boundary is set by two
conditions. The contour nearest the corona marks the boundary between
gas that is bound and gas that is unbound.  The contour nearest the
funnel indicates where the positive radial mass flux has dropped below
a limit we use to define the funnel wall jet, ${(\rho\,U^r)}_{jet}
\gtrsim 10^{-4}{(\rho\,U^r)}_{max}$, where ${(\rho\,U^r)}_{max}$
denotes the maximum {\it unbound} outward mass flux.  The vector field
indicates the direction, but not the magnitude of pressure gradients
and Lorentz forces.

The labels in the figure identify the five main regions that we will
discuss.  These are:
\begin{itemize}

\item[(1)] The {\bf body of the jet} is depicted in Figure
\ref{Jetstills} and consists of the hourglass-shaped (hollow cone)
region where there is significant unbound outward mass flux.  Inside
the cone is the axial funnel where the density if very low, and outside
the cone lies coronal envelope consisting of bound material.

\item[(2)] The {\bf base of the jet} is the origin of the continuous
region of unbound, outgoing radial mass flux. In the high-spin KDP the
base is relatively stable and is marked by a ring of denser gas.  In
low-spin models the base is not so sharply defined, instead showing
both spatial and temporal variability.  

\item[(3)] The {\bf injection region} is where mass is injected into
the unbound outflow.  This region begins where the inner torus, coronal
envelope, and axial funnel converge, at a distance comparable to the
marginally stable orbit, $r_{ms}$.  The mass flux in the jet is not
uniform, but varies with time and increases with increasing radius.

\item[(4)] The {\bf funnel outflow} consists of a tenuous, fast,
unbound outflow.  The axial funnel contains a large-scale poloidal
magnetic field similar to a split monopole, as discussed in Papers I
and II. This field is established during the ejection of low-density
magnetized material into the funnel in the early stages of accretion.
If the black hole is rotating, this funnel field is wound up and
amplified by frame dragging, creating a toroidal field and an
outward-going energy flux (e.g., figs.~6 and 7 in Paper II).

\item[(5)] To discuss possible asymptotic properties of the jet, we
refer to the {\bf outflow region}, which is the distant region well
away from the disk and the black hole where the jet is moving
ballistically.  We take the outflow region to lie just inside the outer
grid boundary which is located at $r=120M$.  Various quantities can be
computed here and compared between the different KD models.

\end{itemize}

In the following subsections, we consider each of these regions in turn.

\subsection{The body of the jet}\label{body}

\begin{figure}[ht]
    \epsscale{1.0}
    \plottwo{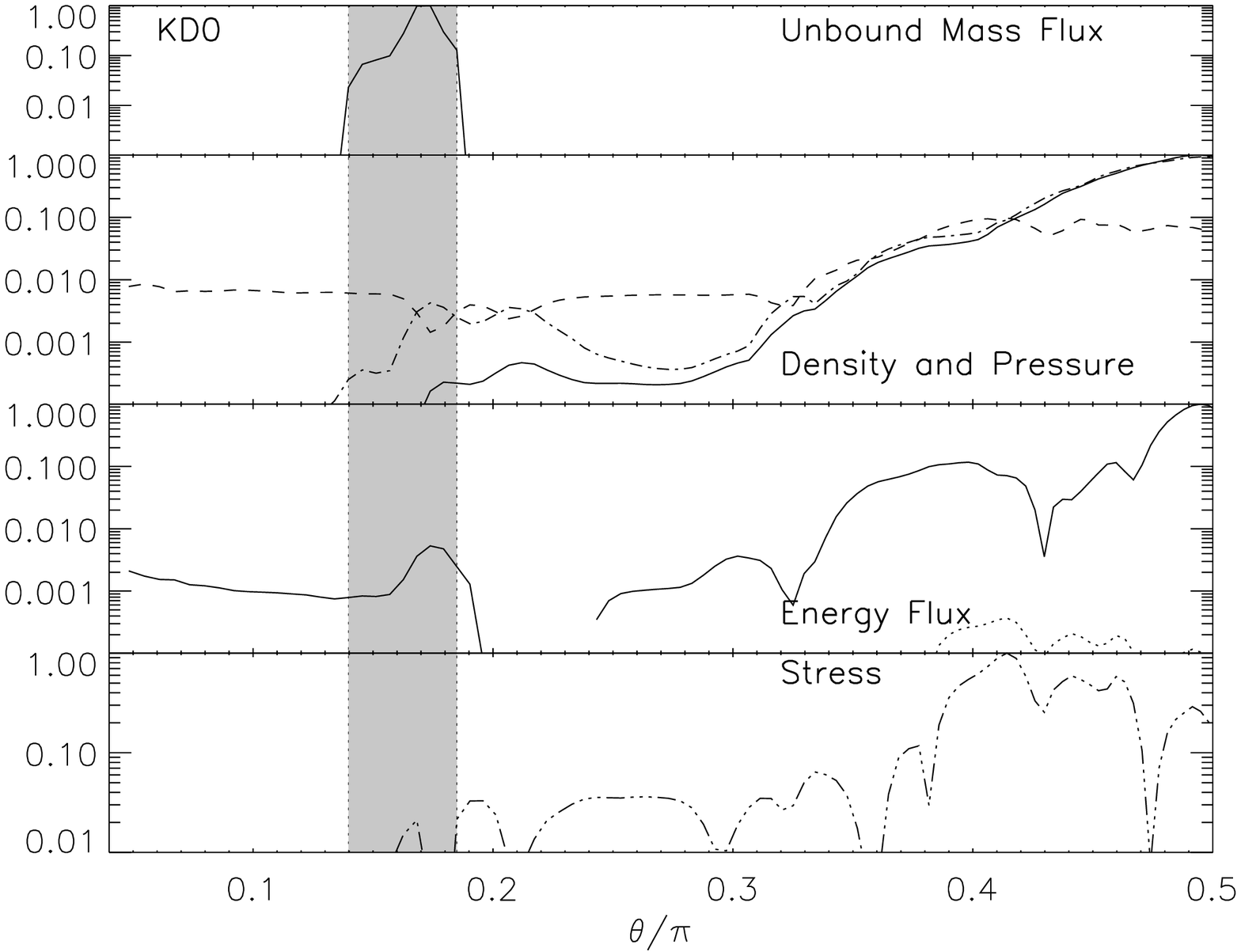}{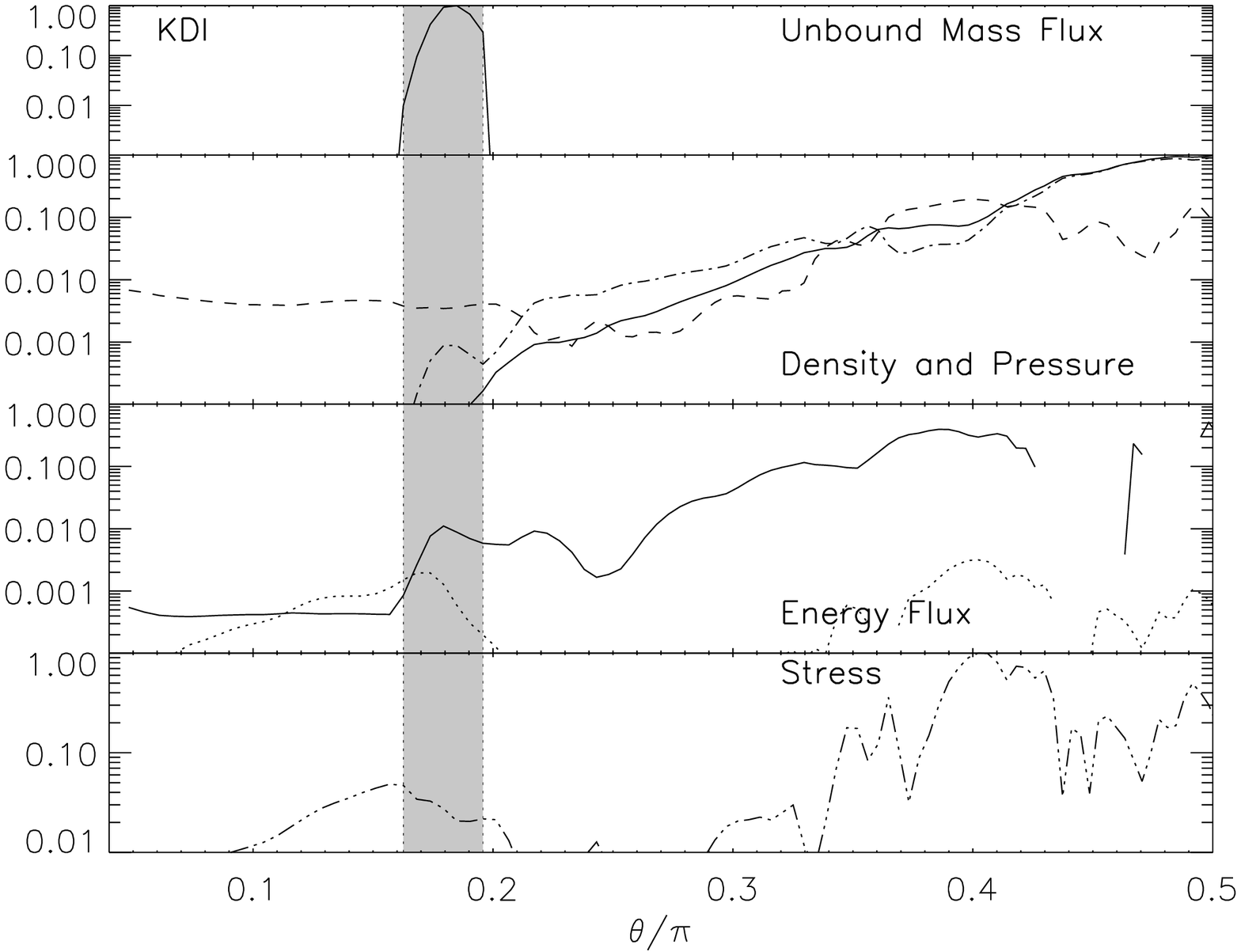} 
    \plottwo{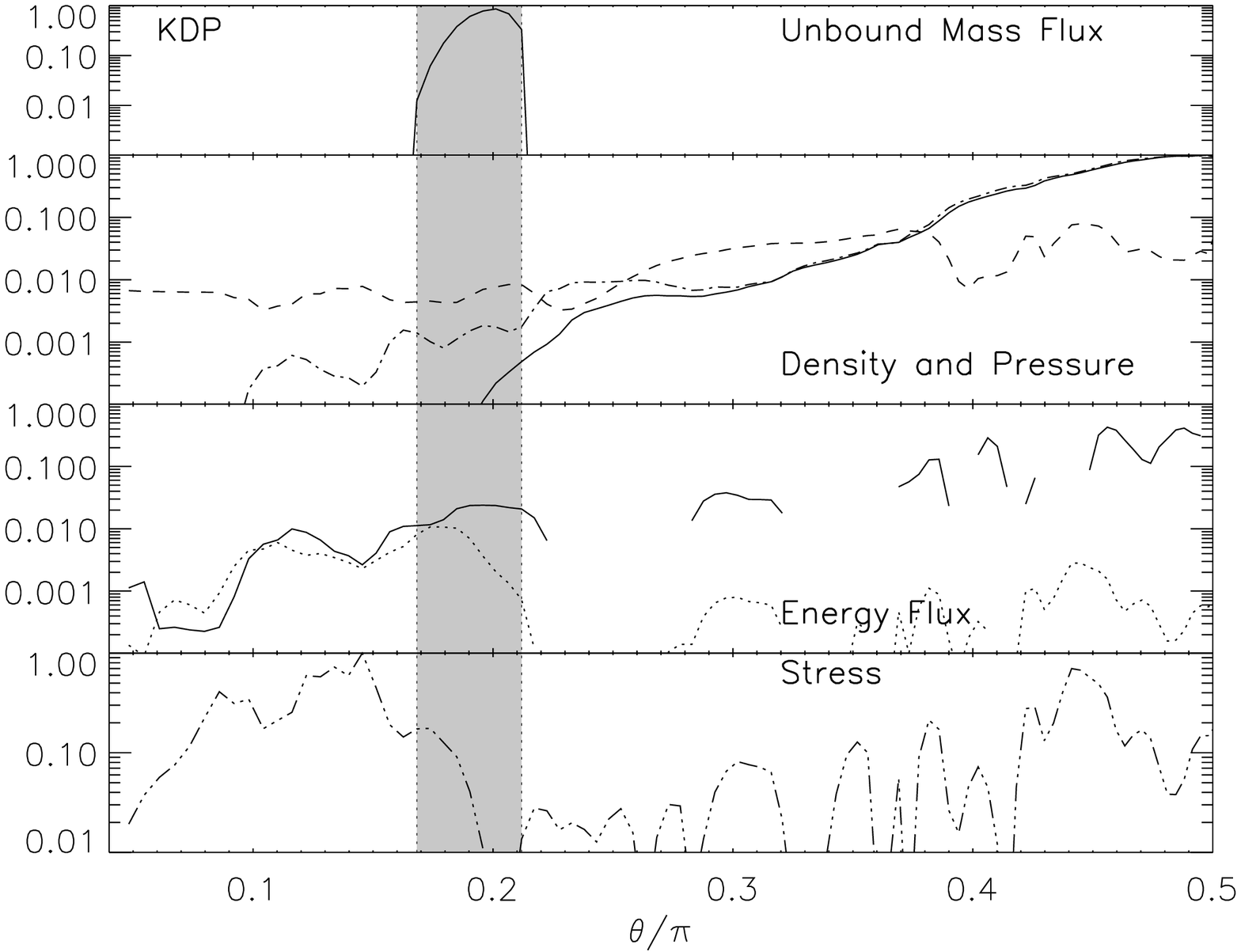}{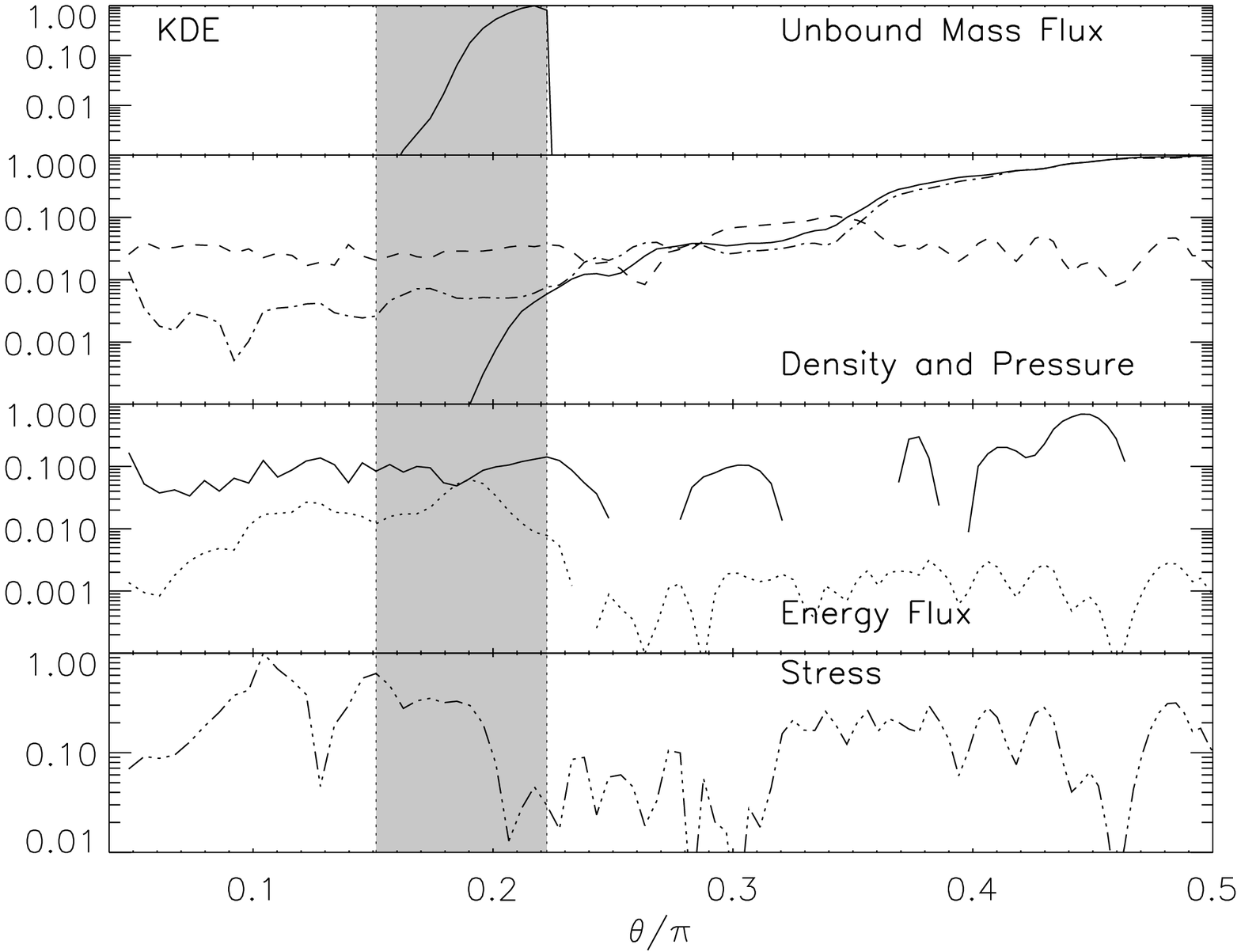}
    \caption{\label{vars} 
    Cross-section of key code variables through the body of the jet
    taken from the late-time simulation data ($t = 8080\,M$) for the
    four KD models. Each plot shows azimuthally-averaged quantities
    plotted as functions of polar angle $\theta$. For consistency across
    models, the cross-sections are taken at a radial distance of
    $10\,r_{ms}$. The top graphs in each panel show mass flux,
    $\rho\,U^r$, which defines the jet boundaries. The second row of
    graphs show density (solid line) as well as gas (dash-dot line) and
    magnetic pressure (dashed line). The third row of graphs shows the
    fluid (solid line) and electromagnetic (dashed line) 
    components of outward energy flux (gaps in the curves correspond to
    radially inward flux). The bottom graphs show the
    absolute value of the magnetic stress.
  } 
\end{figure}

Figure \ref{vars} presents a cut through a representative region of the
jet body for all four KD models.  It depicts four sets of variables as
functions of polar angle, $\theta$, from the ``North'' polar axis to the 
equator; the variables are azimuthally-averaged, and taken at a radial 
distance of $10\,r_{ms}$.

The top graph in each panel is the unbound outward mass flux,
$\rho\,U^r$, as defined above.  The units are the fraction of the
maximum mass flux in the jet.  This defines the funnel wall jet, and
its angular width is represented by the shaded region which extends to
the lower graphs.  One can see that the peak flux lies just inside the
boundary with the corona, and the flux drops off rapidly toward the
funnel.  This peak tends to shift more toward the coronal boundary of
the jet with increasing black hole spin.

The second graph from the top shows gas density, and the gas and
magnetic pressures.  Density is normalized to its peak value, which is
found at the equator.  Gas and magnetic pressure are normalized to the
peak {\it total} pressure, also found at the equator. The density in
all models decreases steadily through the corona, and shows a sharp
roll-off through the jet.  The gas pressure has a similar profile,
except that the decrease through the jet is not as pronounced,
especially in models KDP and KDE.  In the latter case, the gas pressure
through the jet and funnel is roughly constant.  The gas temperature is
quite high in the funnel since the low-density gas has been heated
by shocks driven into the funnel from denser regions of the flow.  The
$\theta$-profile of the magnetic pressure is much more shallow than
that of gas pressure in all the models.  In model KD0, magnetic
pressure is roughly constant through the funnel, dips slightly through
the jet and corona, then rises slightly through the body of the disk.
In models KDI, KDP, and KDE, magnetic pressure is roughly constant in
the funnel and jet, dipping slightly in the corona, near the jet
boundary, before rising again as one moves toward the equator.  As was
pointed out in Paper I (see fig.~8), the total pressure is relatively
smooth through the corona into the funnel.  The funnel is magnetically
dominated, the main disk is gas-pressure dominated, and the corona has
a ratio of pressures near unity.  The unbound mass outflow is found
just inside the region where $\beta = P_{gas}/P_{mag}$ drops below
unity.

The third graph from the top shows the fluid and electromagnetic
contributions to outward energy flux (${T^r}_{t\,{\rm (fluid)}}$ and
${T^r}_{t\,{\rm (EM)}}$), normalized to the maximum of the absolute
total energy flux, which is found near the equator in all cases.  The
fluid component varies in sign in the main disk and corona, as would be
expected for turbulent motions.  Its absolute value shows a steady drop
from the main disk body through the corona.  In the unbound portions of
the outflow, energy flux is everywhere outward.  The fluid energy flux
is roughly constant within the axial funnel; it is due to high enthalpy
and high outflow velocities.  The electromagnetic component of the
energy flux is extremely weak in model KD0, and becomes more prominent
in the funnel with increasing black hole spin.

In the lower graph we plot the absolute value of the magnetic stress,
$\|b^r\,b_\phi\|$, again normalized to its maximum value on the slice
of constant radius.  This quantity exhibits a strong dependence on
black hole spin in the jet and the funnel.  In model KD0, the stress
drops sharply in crossing from the main disk body to the corona. Though
stress is variable through the corona, the peaks are roughly level.
Peak stress in the jet is slightly weaker than in the corona, and
extremely weak in the funnel. The models with a rotating hole also show
a drop in stress in the corona, relative to the main disk. However,
stress in the jet and funnel becomes progressively more dominant with
increasing spin, with the maximum stress found inside the funnel in
models KDP and KDE.  In these three models, stress in the jet drops
from the funnel to the corona side, indicating that the stress acting
on the jet material is strongest where the density is low.

The fluid velocity also makes a sharp transition through the jet, from
predominantly orbital motion in the corona to radial or helical motion
in the funnel.  As noted in Paper II the magnetic field lines also
undergo a corresponding transition through the funnel wall.  The
outward velocity increases sharply through the funnel wall.  The
mass-weighted mean of $V^r$ within the funnel wall jet ranges from
$\approx 0.3 c$ for model KD0 to $0.5 c$ for model KDE.  The velocity
is predominantly radial on the funnel side and toroidal on the corona
side of the jet.  The values are comparable to the toroidal velocities
found in the inner torus which range from $\approx 0.4 c$ in KD0 to
$0.6 c$ in KDE.  The outward speed at a fixed angle is established
quite close in:  the flow accelerates from near its base to 10--$20M$
(the end of the acceleration zone moves outward with decreasing black
hole spin) and then retains that speed all the way to the outer
boundary at $120M$.

\subsection{The base of the jet}\label{base}

We define the base of the jet as the point closest to the black hole
where there is a continuous flow of unbound mass to the outer
boundary.  In general this point varies in time, but model KDP has a
base that is compact, largely axisymmetric, and found at a more or less
fixed location over extended periods of time, making it ideal for
illustration purposes.  Figure \ref{fluxdetail} shows the
azimuthally-averaged radial mass flux in KDP in the inner region of the
flow.  The yellow and red colors indicate outflow, and the circle
locates the base. (Note a small disconnected segment of positive flux
nearer the black hole, which is not part of the jet.) The jet
originates on the surface of the inner torus in the vicinity of
$r_{ms}$ ($=2.32 M$ in KDP).  In this and all models where jets have
been observed, the base of the jet is found where inner torus, coronal
envelope, and axial funnel intersect.

Figure \ref{fluxdetail} also depicts a set of three equipotentials for
marginally bound material that bracket the critical contour ($\Phi_{mb}
= \left[-0.01,0,0.01\right]$, where $\Phi_{mb} =
\log{\left[-U_t(l_{mb},a;r,\theta)\right]}$.  These equipotentials are
of interest for comparison to constant-angular momentum thick disk
models (e.g., Abramowicz et al. 1978).  The base of the jet lies above
the critical contour ($\Phi_{mb} = 0$), which suggests that the
outbound flow of the jet lies up against the centrifugal barrier.  The
jet appears to be squeezed by the pressure of the inner torus and
corona against the centrifugal barrier that defines the funnel.  Since
material with angular momentum cannot encroach into the funnel, the jet
material is compelled to exit along the funnel wall.

The launching of a jet appears to be facilitated by a geometrically
thick inner torus deep in the potential well of the black hole.  As
discussed in Paper I, the inner torus moves closer to the black hole
and becomes denser and thicker with increasing black hole spin.  This
spin-dependent increase in the size of the inner torus provides one
indirect way in which black hole spin could increase the mass flux in
the jet.  Both the jet mass flux and the accretion rate through the
inner torus are highly time variable.  In animations of the funnel wall
jet, increases in jet flux appear to be correlated with changes in the
inner torus, suggesting that a build up of the inner torus can be one
way to increase the mass flux through the jet.

\subsection{The injection region}\label{injection}

\begin{figure}[ht]
    \epsscale{1.0}
    \plotone{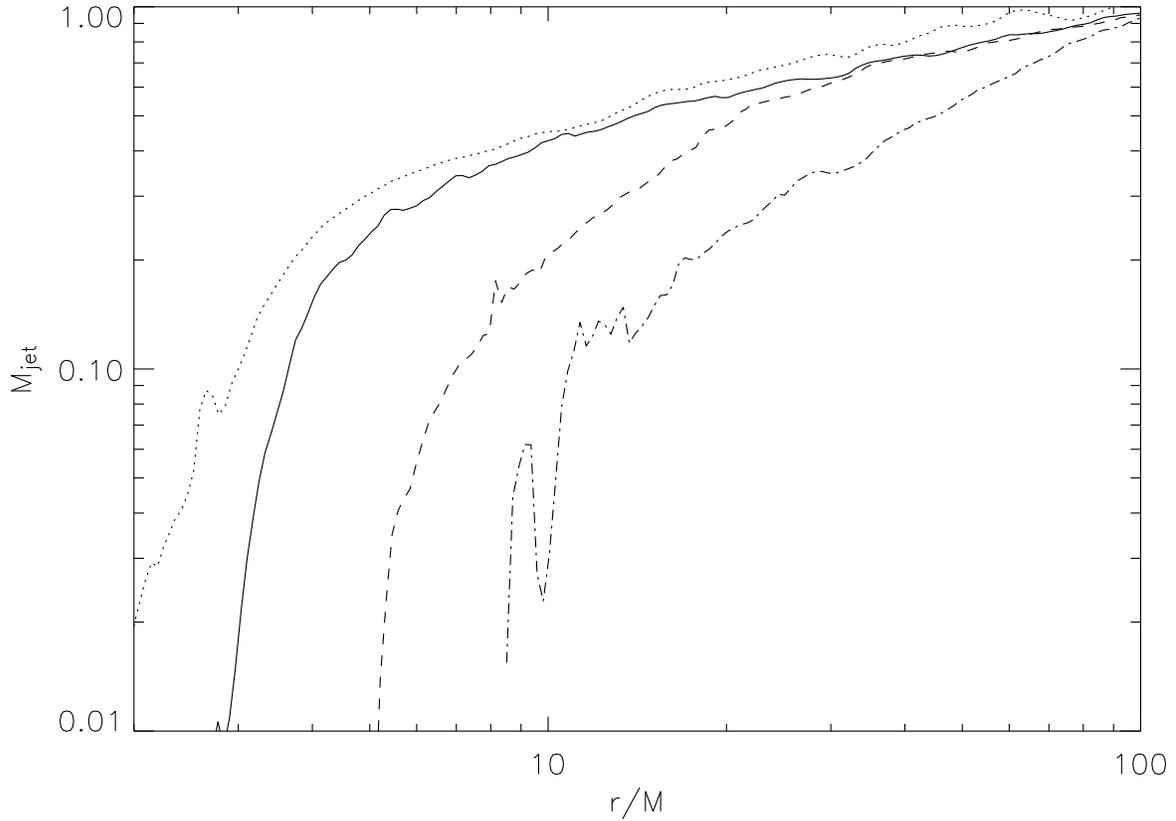} 
    \caption{\label{jetmass} 
    Mass distribution in the jet, $M_{jet}(r) = \{\rho\,U^r\}_{jet}$,
    for model KDP (solid line), KD0 (dash-dot line), KDI (dashed line)
    and KDE (dotted line).  Each curve is normalized to the maximum,
    which occurs at the outer radial boundary. The shallow slope at
    large radii indicates that mass entrainment is taking place in
    this region for all models.  The steep slope at small radii
    indicates the rapid injection of mass into the jet over a
    small radial distance.  This injection region is mostly
    missing in model KD0, and strongest in the high-spin
    models KDP and KDE.
  } 
\end{figure}

\begin{figure}[ht]
    \epsscale{1.0}
    \plotone{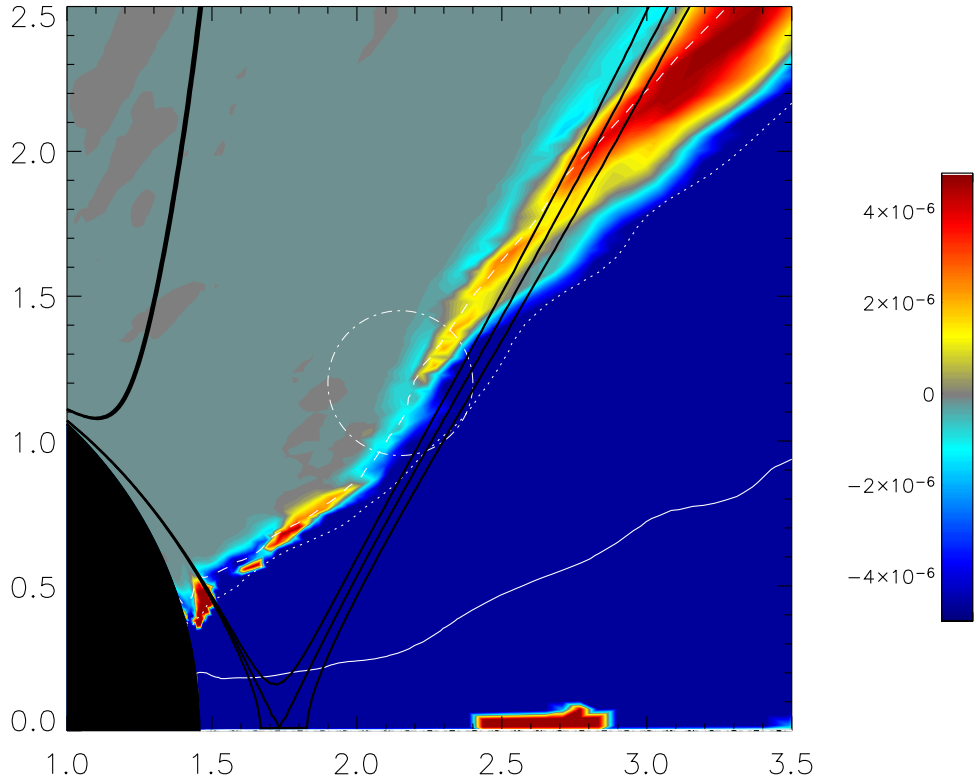} 
    \caption{\label{fluxdetail} 
    The azimuthally-averaged radial mass flux, $\langle
    \rho\,U^r\rangle_\phi$, near the base of the jet for model KDP (at
    $t = 8080\,M$), above the equatorial plane. Three density contours
    have been added to locate the disk, $10^{-3}\, \rho_{max}$ (dashed
    line), $10^{-2}\, \rho_{max}$ (dotted line),         $10^{-1}\,
    \rho_{max}$ (solid line). The base, highlighted by a circle,
    features both an outgoing and ingoing component. The extended
    region of outbound material outside the $10^{-3}$ density contour
    is unbound (this density contour approximately follows the funnel
    wall in this region of the flow). Equipotentials for marginally
    bound gas (see text for details) are also shown as thick black
    lines; the contours bracket the critical contour, $\Phi_{mb} =
    \left[-0.01,0,0.01\right]$. } 
\end{figure}

The injection region is where mass is transferred into the unbound
outflow.  Figure \ref{jetmass} shows the mass distribution as a
function of radius in the jet, $M_{jet} (r) = \{\rho\,U^r\}_{jet}$, for
all four KD models.  These curves show two features:  a region of
relatively steep slope near the black hole, and a more extended region
of shallow slope at larger radii.  This extended region shows that mass
entrainment takes place along the jet/corona boundary (if all matter
injection were taking place near the base, this portion of the curve
would be flat).  The region of steep slope at small radii indicates
that rapid matter injection occurs near the base of the jet.  The point
where mass injection begins moves inward with increasing black hole
spin.  The slope of the curve also steepens with increasing black hole
spin.  Apparently mass-loading of the jet can be enhanced
by effects due to black hole spin.

In Paper I we noted that the mass flux in the jet increases outward
along the radial extent of the inner torus, implying that the jet
material found near the inner torus originates in it.  The specific
angular momentum, $l$, provides one marker to indicate the origin of
the jet material, and near the black hole the specific angular momentum
in the jet is comparable to the specific angular momentum in the
adjacent inner torus at the same radius.   Of course the centrifugal
barrier ensures that $l$ undergoes a sharp poleward transition through
this region; high-$l$ material is not found in the funnel.  The thin
layer of corona between the inner torus and the injection region
contains elevated poleward mass flux ($\rho\,U^\theta$), also
suggesting that matter is moving from the inner torus to the corona,
and hence to the jet.  This poleward mass flux cuts off abruptly at the
coronal interface to the jet, exactly where the region of strong
outward radial mass flux begins.

\begin{figure}[ht]
    \epsscale{1.0}
    \plotone{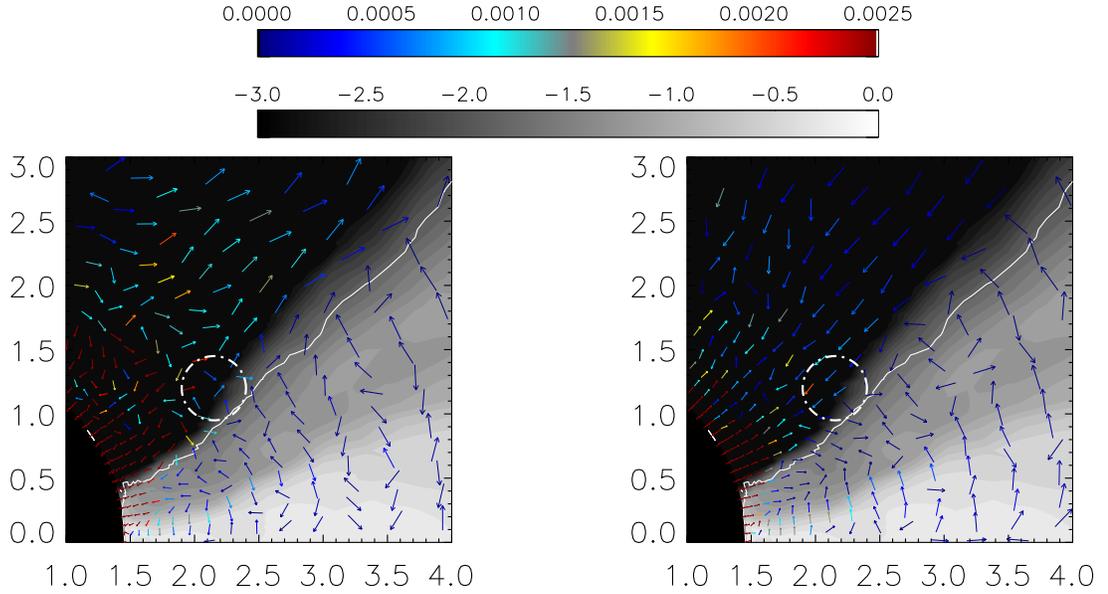} 
    \caption{\label{KDPforce} 
    The azimuthally-averaged poloidal components of the Lorentz force
    (left panel) and pressure gradient (right panel) near the base of
    the jet, model KDP (at $t = 8080\,M$). The vectors carry magnitude
    information in their color (see accompanying scale, in code units);
    the length of the vectors is proportional to the local zone width.
    For reference, azimuthally-averaged density contours are shown on
    grey-scale. The base of the jet is highlighted by a circle, and
    the solid white line marks the $-h\,U_t = 1$ contour, which marks
    the coronal boundary of the funnel-wall jet.  Near the base, the
    Lorentz and pressure gradient contributions to the force are normal
    to the surface of the disk and divert matter from the inflow into
    the jet.
    } 
\end{figure}

What forces are responsible for injecting mass into the jet near
the inner torus?   Figure \ref{KDPforce} shows vector plots of the
poloidal components of the Lorentz and pressure-gradient forces
[eqns. (\ref{pequation}) and (\ref{lfequation})]. The
plot region is the same as that of Figure \ref{fluxdetail}, 
and the circle indicates the location of the base of the jet.
The arrows in the vector plots are scaled to the local zone dimensions
and oriented along the local field direction.  The color of the arrows
is set by the magnitude of the force (in code units), and
can be read on the color scale.  The $-h\,U_t = 1$ contour, 
delineating the unbound flow, is shown as a thick white line.  In
addition, the azimuthally-averaged contours of density in the inner
torus have been added in grey-scale.

   This figure shows that in the inner torus and adjacent corona,
the orientation of the poloidal force
component is largely perpendicular to isodensity contours, consistent
with pressure-supported hydrostatic equilibrium.  In
this figure the magnitude of the poloidal components of the gas
pressure-gradient and Lorentz force are comparable, but since the inner
torus and the thin corona surrounding it are highly dynamical, this is
not always true.  At other times there are regions within the corona
where one or the other force components is dominant (see, e.g., fig. 8,
Paper I which shows $\beta$ in the disk and surrounding corona).
Beyond the base of the jet,
the net poloidal force is radially outward, with the outward-directed
Lorentz force dominating the inward-directed gas pressure gradient. As
noted earlier, the Lorentz force as shown here combines contributions
from magnetic pressure gradients and stress terms. In the inner torus,
it is the magnetic pressure gradient that contributes most to the
Lorentz force; in the injection region and jet body, it is the stress
contribution that is significant.

\begin{figure}[ht]
    \epsscale{1.0}
    \plotone{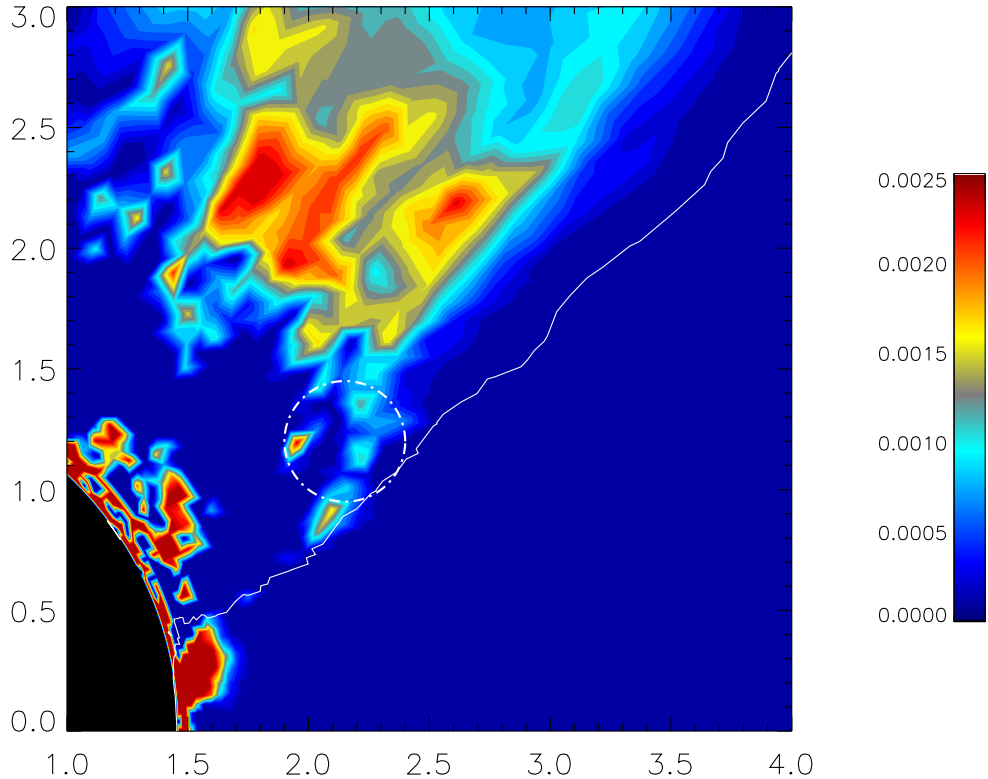} 
    \caption{\label{KDPforcetor} 
    The azimuthally-averaged toroidal component of the Lorentz force
    near the base of the jet, model KDP (at $t = 8080\,M$). The
    magnitude of the force is indicated in the accompanying color scale
    (in code units). The base of the jet is indicated by a circle,
    and the solid white line is the $-h\,U_t = 1$ contour, which
    marks the boundary of the unbound outflow. The Lorentz
    force shows a strong contribution both at the base of the jet, and
    on the funnel interface of the jet. A similar plot of the toroidal
    component of the pressure gradient (not shown) indicates that the
    Lorentz force is dominant in the funnel.
  } 
\end{figure}

Figure \ref{KDPforcetor} shows the azimuthally-averaged
\textit{toroidal} component of the Lorentz force near the base of the
jet for model KDP.  The plot region is the same as that of Figure
\ref{KDPforce}, as is the color scale for the magnitude of the force
component. This component of the Lorentz force corresponds to the
familiar $B_r\nabla_r B_\phi$ Maxwell stress in nonrelativistic MHD;
the toroidal gas and magnetic pressure gradients are negligible.  The
strength of the Maxwell stress is correlated with the spin of the black
hole: it is weakest in model KD0, and strongest in model KDE.  The
Maxwell stress is prominent through the base of the jet, and passes
through the base of the jet to the funnel side of the of the jet in the
injection region and the jet body beyond.  This pattern is seen in all
models, and is indicative of magneto-centrifugal launching of the jet.

\subsection{The funnel outflow}\label{funnel}

The general properties of the funnel outflow were discussed in Paper I,
and its magnetic field in Paper II. Hydrostatic equilibrium is not
possible within the funnel:  material with significant specific angular
momentum cannot get into the funnel, and any low-$l$ material must
either accrete into the hole, or leave as an outflow.  Hence, the
funnel remains largely evacuated throughout the simulations for all
models; indeed, the density in the funnel approaches the floor value in
some locations, particularly at large radius.  Outflow velocities in
the funnel are high, with radial values in the range $V^r \simeq
0.8$--0.95; these numbers should be regarded as qualitative indicators
only.

In the present context, the most important aspect of the funnel is the
presence of a large-scale organized poloidal magnetic field.  Paper II
showed that the funnel field has a radial structure similar to a
split-monopole.  This field is created as a result of wholesale field
expansion:  some of the closed fieldlines initially contained within
the dense accretion flow near the equator slip outward into the region
of very low mass density when the mass initially attached to them
drains into the black hole.  As they do so, they expand outward,
creating the large-scale poloidal field that occupies the axial
funnel.  If the black hole is rotating, the funnel field is wound up
and the toroidal field is amplified by frame dragging. Once the field
in the funnel becomes comparable in magnitude to the adjacent field in
the corona, its strength saturates and little further flux is drawn
into the funnel.

As Figure \ref{vars} shows, the funnel outflow is a region of elevated
outward energy flux but extremely weak mass flux.  Although the density
in the funnel outflow is very low, it has high internal energy.
Compressive modes in the disk's turbulence become nonlinear as they move
down the steep density gradient at the edge of the funnel. Entropy is
created in the resulting shocks, and the very low-density material in
the funnel interior rises to very high temperature as a consequence.
This leads to an enthalpy-dominated fluid energy flux.  The
electromagnetic energy flux increases with black hole spin as toroidal
field is generated by frame-dragging.

\subsection{The outflow region}\label{outflow}

In Paper I we noted an increase in the prominence of the jet's radial
mass flux, in relation to typical values in the main disk body, with
increased black hole spin (e.g., fig. 10, Paper I).  To put this on a
more quantitative footing, we will examine the ejection rates of mass,
energy, and angular momentum for unbound and bound material in the outflow
region, taken to be just inside the outer radial boundary of the
computational domain.  The fluxes are computed from the periodic full
data sets which allow the values for bound and unbound flows to be
distinguished.

\begin{figure}[ht]
    \epsscale{1.0}
    \plotone{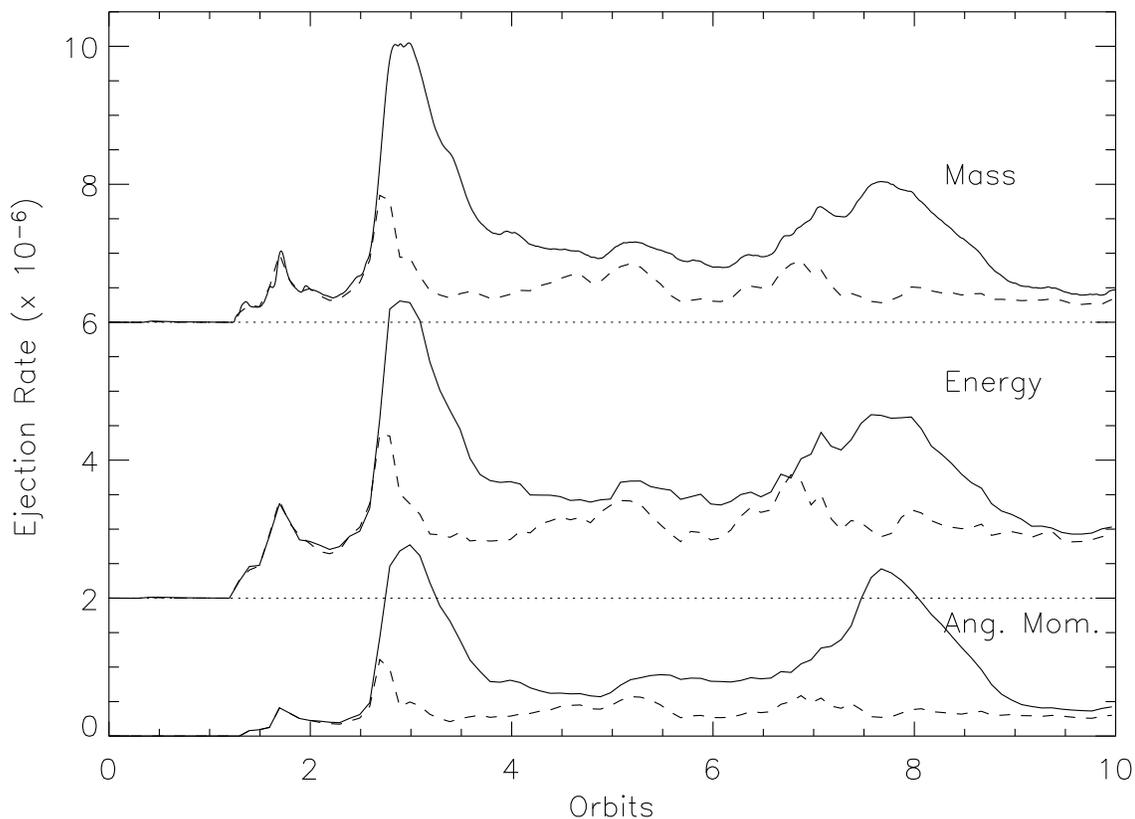} 
    \caption{\label{KDP_Ejection} 
    The ejection rate of mass, energy, and
    angular momentum through the outer radial boundary as a function of
    time for model KDP. The solid lines represent the rate for both
    bound and unbound matter; the dashed lines represent the unbound
    portion.  The units are fraction of initial torus
    mass/energy/angular momentum accreted per unit time $M$. For
    clarity, the mass and energy curves are shifted vertically by 7.0
    and 3.0 $\times {10}^{-6}$, respectively.
  } 
\end{figure}

Figure \ref{KDP_Ejection} shows a time-history of the ejection rates
for mass, energy, and angular momentum for model KDP over the full 10
orbits of the simulation; the other KD models are similar. The solid
curves show the total ejection rate, which contains contributions from
the funnel outflow, the jets, and the bound coronal outflow.  The
specific value for the bound outflow is, of course, dependent on the
location of outer grid boundary.  The dashed lines show the ejection
rates for the unbound portion flow which, because it is unbound, is
expected to be largely independent of the location of the outer
boundary.

Between times $t \approx 1.0\,T_{orb}$ and 3.5 $T_{orb}$ there are a
weak and then a strong peak in the flux.  The first peak is almost
entirely made up of unbound material blown off the initial torus as the
evolution begins; it reaches the outer boundary at about 1.5 orbits of
time.  The second, larger flux peak is made up of both bound and
unbound material associated with the initial accretion into the black
hole.  As a thin equatorial stream of matter nears the black hole some
material splashes back.  Part of this outflow propagates along the
funnel-wall, a precursor to the more permanent jet, and part of it
comes from a bound outflow that moves along the surface of the disk
(this surface outflow was noted in the SF models of DH03b).  This mix
of bound and unbound material takes about one orbital period after
accretion begins to reach the outer boundary, giving rise to the second
spike (this corresponds to a non-relativistic velocity $v \sim
r_{max}/T_{orb}=120\,M/800\,M \sim 0.15\,c$).  As the accretion flow
becomes more fully established, the outflowing corona begins to
contribute to the bound ejection rate.  The jet emerges as a stable
structure, with a continuous ejection of unbound material.  The
ejection rate remains somewhat variable, on a time scale comparable to
the orbital period of the main disk, in contrast to the rapid
variability of the accretion rate into the black hole (fig.  14 of
Paper I).  After the initial transients the unbound ejection rates have
a mean value that is $\sim 10^{-6}$ of the initial mass/energy/angular
momentum of the torus per unit time $M$ for all models.  This rate can
be compared to the mean accretion rate into the black hole as reported
in Paper I; these range from $1.78\times 10^{-5}$ for KD0, decreasing
to $4.3\times 10^{-6}$ for KDE.

Table \ref{Eject_fraction} lists the ejection rates for mass, energy,
and angular momentum, computed using equation (\ref{surfflux}).  The
integrals were taken over the full simulation for each model in order
to provide a consistent basis for comparison.  The table lists the KD
models, as well as other models that will be discussed in \S
\ref{other}. For each quantity ($Q$), five columns are shown: the
initial value, $Q_0$; the accretion rate of bound material through the
inner boundary, $\Delta Q_{\rm i}$; the bound (coronal) ejection rate,
$\Delta Q_{\rm b}$; the unbound (jet and funnel outflow) ejection rate,
$\Delta Q_{\rm u}$; and finally, the efficiency, $\eta_{\rm Q}=\Delta
Q_{\rm u}$/$\Delta Q_{\rm i}$, expressed as a percentage.  For the KD
models, the ratio of unbound ejection to accretion increases with
increasing spin, as can be seen by comparing $\Delta Q_{\rm u}$ to
$\Delta Q_{\rm i}$.  The unbound mass flux normalized to the initial
mass increases with black hole spin by a factor of 10 from KD0 to KDE.
There is a similar increase in both the energy and angular momentum
fluxes in the jets relative to the accreted quantities as a function of
black hole spin.  This suggests that the black hole spin itself may be
a source of energy for the jets.

The efficiency measure, $\eta_{\rm Q}$, compares ejection through the
unbound outflow to accretion onto the black hole.  The data in Table
\ref{Eject_fraction} shows a spin-dependent increase in this value
for mass, energy, and angular momentum.  Furthermore, there is a
substantial jump in efficiency between models KDP ($a=0.9$) and KDE
($a=0.998$), especially for angular momentum. This jump is a clear
manifestation of the non-linear contribution of black hole spin to the
dynamics, and a demonstration that near-extreme Kerr holes can
generate highly efficient jets.

In the KD simulations the initial torus serves as the mass source for
the accretion flow.  The transport of angular momentum allows accretion
to occur, forming an accretion disk that extends down to near the black
hole, where an inner torus forms near $1.5 r_{ms}$.  In following the
mass flow through this system there are four quantities of interest:
the bound flux through the radial shell at $r=15.0\,M$ (the inner edge
of the initial torus), the amount of bound mass accreted into the black
hole, and the amount of unbound mass carried off by the jets.  What
remains after accounting for these fluxes should be equal to the mass
contained in the region inside $15M$ at the end of the simulation.
Table \ref{mass_table} provides the numbers computed from the data
dumps, and expressed as fractions of the initial torus mass for each
model.  The bound mass accreted into the black hole ($\Delta M_{hole}$)
and unbound mass carried off by the jets ($\Delta M_{jet}$) are taken
from Table \ref{Eject_fraction}. The flux of bound matter in the main
disk and corona through the shell at $r=15.0\,M$ ($\Delta M_{r=15}$),
is computed using the mass flux integral, $\langle \rho\,U^r\rangle_t
(r=15.0)$ with the condition $-h\,U^t < 1$. The final mass is computed
at the end of the simulation by direct integration of the gas density
inside $r = 15.0\,M$.  $M_{\rm final}$ should equal the amount that
passed through $r=15M$ minus the losses to the hole and in the jet.
The agreement in these numbers is good, but not perfect.  Two 
sources of error in this comparison are the possible entrainment of
mass into the jet outside of $r=15M$ in all models, and the need to
evaluate the mass fluxes using the data dumps which are spaced at
intervals of $80M$.  One conclusion, however, is that unbound mass flux
in the jets is a significant component of the mass flow through the
near hole region for high-spin holes.  The accretion flow at $r=15M$ is
mainly established by the MHD turbulence in the main body of the disk.
The accretion rate into the hole, and, to some degree, the outflow rate
in the jets are determined by processes near the black hole, and those
rates need not match the accretion rate supplied by the main disk.  One
consequence is the spin-dependent build-up of the inner torus as noted
in Paper I.

\section{Jets in other simulations}\label{other}

In the preceding section we examined the unbound outflows that arise in
the KD simulations.  Jets have been seen in other accretion simulations
as well.  The present results raise several interesting questions that
can be studied further by an examination of jets in those other
simulations.

\subsection{Jets in the SF simulations}

The SF models (DH03b) consist of initial constant-$l$ tori containing
weak poloidal field loops.  Three black hole spin values were
considered:  zero spin (SF0), prograde $a/M = 0.9$ (SFP), and
retrograde $a/M = -0.998$ (SFR).  The main difference between the SF
and KD models is that the initial torus is hotter in the SF models, and
the SFR model represents an example of a counter-rotating black hole.
We find that jets are produced in all three SF models, including the
extreme retrograde model SFR.  This is noteworthy since the marginally
stable orbit, at $r=9\,M$, is quite distant from the black hole, and
one might expect that the thin, elongated plunging inflow would not be
conducive to jet formation.  The base of the jet in the SFR model
occurs just outside the static limit, i.e., well inside the marginally
stable orbit.   Despite this contrast, a jet is still produced in the
retrograde case, although it is considerably weakened compared to the
prograde models.  The funnel wall is more flared out in SFR than in the
prograde simulations, and consequently the jet has a relatively large
opening angle near its base.

Table \ref{Eject_fraction} presents information on the ejection rates
for the SF models. It was noted in DH03b that these disk models
generate rapid accretion, a fact that is reflected in the elevated
rates shown in the table, e.g. 34\% of the initial torus mass is
accreted for model SF0, whereas only 16\% is for model KD0. The greater
accretion rate in the SF models is most likely connected to their
greater heat content and therefore greater thickness. To the extent
that magnetic stresses grow as longer-wavelength magneto-rotational
modes become unstable, greater disk thickness promotes faster
accretion.  The ejection rates in the SF models are also elevated.
This could arise from at least two possible causes: the greater rate at
which mass is fed into the inner regions of the accretion flow, making
it available to be expelled in the jet, and the greater amount of
energy released by accretion, some of which can be utilized for driving
outflows. One can regard energy drawn from the rotation of the black
hole as catalyzed by accretion, so that the rate of rotational energy
extraction, too, is proportional to the accretion rate.

\subsection{Jets in pseudo-Newtonian simulations}

There have been several investigations of black hole accretion in the
pseudo-Newtonian approximation (Hawley \& Krolik 2001, 2002; HB02; Kato
{\it et al.} 2004). The initially poloidal simulation of Hawley \&
Krolik (2002) is very close in character to the KD0 model because it
began from a hydrostatic initial condition with pressure maximum at $r
= 20M$ and a purely poloidal magnetic field, and the pseudo-Newtonian
potential was designed to mimic the principal properties of the
Schwarzschild metric. HB02 was similar in design, but differed in that
the initial condition was a constant angular momentum torus centered at
$r = 200M$. The work of Kato {\it et al.} (2004) was similar to that of
HB02, but for two different initial field strengths, and non-zero
initial external gas pressure surrounding the accretion disk.  Like
HB02, the accreting matter was centered at relatively large distance,
in this case $80M$.

In Hawley \& Krolik (2002), although a high-temperature and (sometimes)
magnetically dominated region is formed within a conical region near
the axis, no true outflow is created; all the matter remains bound. By
contrast, in HB02 a structure similar to that of KD0 was seen, a
narrow outflow of unbound gas moving outward along the boundary between
the corona and the funnel. At fixed height above the equatorial plane,
the density is roughly constant in the corona, but then drops rapidly
toward the axis. Gas pressure also drops through the jet body, but
levels off in the funnel. In the corona, magnetic pressure dominates,
but, unlike the KD models, the magnetic field in the funnel is not
strong and $\beta$ remains greater than 1. The field strength in the
funnel has likely been reduced by the axial boundary condition, but
this seems to have had little overall effect on the flow, which
resembles what is seen in KD0. The primary jet acceleration mechanism
in the HB02 model is the high gas pressure found near the centrifugal
barrier. Jet collimation is due to the magnetized corona surrounding
the jet. Kato {\it et al.} (2004) find a magnetic field configuration
similar to the one reported in Paper II for the Schwarzschild hole
simulation: radial field in the funnel surrounded by predominantly
toroidal field. The jet that develops is driven by strong toroidal
field pressures, and propagates in a $\beta \sim 1$ environment.  The
jet is collimated in large part by pressure from the surrounding
corona.

Although pseudo-Newtonian dynamics can, in many respects, be a good
approximation to the dynamics in a Schwarzschild metric, the effects of
black hole rotation cannot be reproduced in this way.  The dramatic
strengthening of the outflow with increasing $a/M$ is an intrinsically
relativistic effect.

\subsection{$\Gamma=4/3$ equation of state}

Since gas pressure acceleration plays a role in the acceleration of
the jet, it is of interest to see how the jet might vary with the
adiabatic index, $\Gamma$.  Model KDPi is similar to KDP except that
it employs the softer $\Gamma = 4/3$ equation of state. From Table
\ref{Eject_fraction} we see that this model yields a slightly weaker
jet than the corresponding $\Gamma = 5/3$ simulation; only 0.21\% of the
initial torus mass is ejected in the jet, compared to 0.30\% for KDP. A
similar reduction occurs for energy and angular momentum.  However,
the accretion rates are similarly affected, resulting in comparable
efficiencies for models KDP and KDPi.  Any dependence of the jet on the
equation of state in this simulation seems to be modest.

\subsection{Initial toroidal field}

In this simulation, the Schwarzschild (KD0) torus is initialized with a
moderate ($\beta = 10$) {\it toroidal} field. The grid resolution is
$128 \times 128 \times 64$. This model eventually generates an
accretion flow produced by MRI-driven turbulence. What is notable for
the present purposes is that this model generates essentially no jet.

Several factors may all contribute to suppressing the jet. The
accretion rate is considerably weaker in this simulation, and the flow
does not form a substantial inner torus, so there is no large pressure
gradient to push material against the funnel wall.  But, perhaps more
importantly, there is no significant poloidal magnetic field in the
funnel. In the KD models, an extended poloidal field is generated by
the plunging accretion flow, and this field is ejected into the funnel
when that flow reaches the black hole. This becomes the radial funnel
field, whose polarity is established by the initial conditions. In the
toroidal field model, there is no initial north/south field mirror
symmetry, and no large-scale, systematic poloidal field lines generated
by the inflow.  Thus the toroidal field model provides an example of
jet formation switched off, while illustrating that radial funnel
fields need not be the inevitable consequence of accretion.

\subsection{Numerical resolution}

The values shown in Table \ref{Eject_fraction} for low-resolution model
KDPlr are comparable to those of model KDP. The amount of material
ejected through the jet is very similar for mass, energy, and angular
momentum. However, the corresponding accreted quantities are
systematically lower, yielding slightly higher efficiencies for model
KDPlr. As noted in Paper I, the increased numerical resolution in the
``high resolution'' KD models better captures the turbulent and highly
dynamical accretion flow. However, the constancy of the ejected rates
between the two models suggests that the present numerical resolution
is adequate to capture the qualitative behavior of the jets.

\subsection{Axisymmetric Simulations}

To the extent that our results reflect physical processes likely to
occur in accretion systems, they should be reproducible and seen in
other global models.  Although global simulations of this type are
still uncommon, jets similar to what we report here have been observed
by McKinney \& Gammie (2004) who have carried axisymmetric GRMHD
simulations of a torus around an $a/M=0.938$ black hole.  They find the
same general late-time flow structures as characterized in Paper I,
including a conical mass outflow along the funnel edge, along with a
significant electromagnetic flux in the funnel.  The unbound material
in this outflow begins just inside the equipartition contour $\beta =
1$.  The funnel is evacuated and contains a large-scale radial field.
Because their simulations make use of a quite distinct numerical
algorithm, and greater (albeit two-dimensional) grid resolution, the
overall agreement of the results is gratifying.

\section{Conclusion}

Unbound outflows are a natural outcome of accretion in our
self-consistent MHD disk simulations. They carry a non-negligible flux
of energy and momentum, and are collimated and confined to the axial
funnel. These outflows are distinct from the coronal outflows which
remain bound. There are two components to the unbound outflows: a hot,
fast, and tenuous outflow through the axial funnel and a colder,
slower, massive jet confined to the funnel wall by the centrifugal
barrier.  We observe jets in a wide range of simulations, for both
rotating and nonrotating black holes.  While the jets appear in quite
general conditions, their strength is determined by the accretion rate,
the pressure in the inner torus, and the spin of the black hole,
factors which are themselves interdependent.  The funnel outflows have
a strong spin-dependent growth in energy flux, which is their
distinguishing characteristic.

Both gas pressure gradients and Lorentz forces in the inner torus seem
to play a significant role in launching the jets.  The base of the jet
is located at a point close to the black hole where the inner torus,
corona, and funnel meet.  Mass is injected into an unbound outflow and
that is driven by magneto-centrifugal acceleration due to the disk
rotation.  This mechanism can operate even around Schwarzschild black
holes, although the jets produced are relatively weak.  Strong gas
pressure gradients can also play a role.  These gradients are the
predominant acceleration mechanism in the pseudo-Newtonian model of
HB02, and funnel-wall jets can even occur in purely hydrodynamic black
hole accretion flows if the inflowing gas is hot enough and there is
pressure confinement in the coronal region (Hawley 1986).

The KD models develop a
large-scale poloidal, mainly radial, magnetic field in the funnel which
comes from magnetic field injected into the funnel when the accretion
flow first reaches the black hole.  The resulting field is in pressure
balance with the surrounding corona. The poloidal field lines extend
into the ergosphere and this permits frame dragging to tap into the
rotational energy of the black hole, increasing the flux of energy in
the funnel outflow and the funnel-wall jet in the models with a
spinning black hole. Although this mechanism is reminiscent of the
scenario proposed by Blandford \& Znajek (1977) in that energy is
extracted by the magnetic field from a rotating hole, the role of
matter is much more important here than in the original
Blandford-Znajek picture.  Much of the energy flux in the funnel
outflow and funnel-wall jet comes from the fluid enthalpy, not just the
electromagnetic component.  In addition, in the classical
Blandford-Znajek picture, the density was so low everywhere that the
field was taken to be ``force-free''; here, by contrast, most of the
mass and energy flow occur in the funnel-wall jet, where the ratio of
magnetic field energy density to matter energy density is generally
less than unity.

In the SF simulations jets are generated by both prograde and
retrograde accretion flows, although the jets are considerably weaker
in the latter case.  The elevated accretion rates of the SF models are
accompanied by higher ejection rates; the thicker and hotter inner tori
in these models may be part of the cause.  We see only a weak
dependence on the equation of state, but the role of the vertical
thickness and temperature of the inner torus in jet formation is a
topic requiring further study.  Pseudo-Newtonian simulations can also
generate jets comparable in strength to those observed in Schwarzschild
simulations.  The apparent differences in jet strength and driving
mechanisms between the various pseudo-Newtonian simulations raise
several questions for further study.  Together, these simulations
provide support for the idea that unbound outflows are a natural
outcome of accretion.  However, it is possible to switch off jet
production if the accretion rate into the inner region of the
computational volume is significantly reduced, or when there is no
significant poloidal field in the funnel, as was the case when an
initially toroidal magnetic field configuration was used.

In regards to more general issues, our results certainly support the
position that MHD is essential for jet formation and collimation.  The
prospects for purely hydrodynamic processes do not look promising.
While gas pressure alone can accelerate gas away from a sufficiently
hot inner torus, any angular momentum that gas has will keep it out of
the funnel, and collimation can occur only if there is a substantial
corona providing confining pressure.  Even in the pseudo-Newtonian
simulation of HB02, where the jet acceleration is pressure driven and
funnel fields are not important, the confining corona is magnetically
dominated.

But do jets require an initial net large-scale poloidal field? While
such fields clearly can produce substantial jets, they are not
essential.  The jets produced in these simulations developed from
initial conditions in which the magnetic field was completely contained
within a gas torus.  Most of the field subsequently generated by the
accretion disk is toroidal, and this field fills the corona and helps
to collimate the jet.  The large-scale poloidal field that forms is
confined mainly to the funnel, and its strength is limited to
equipartition with the surrounding coronal pressure.  However, even
this modest field can have a quantitative effect on the jet as it
facilitates the extraction of spin energy from the hole.

Finally there is the question of what specific physical mechanism
produces jets.  Here we can give no simple answer.  Strong pressure
gradients, both magnetic and hydrodynamic, magnetic forces analogous to
those considered by Blandford \& Payne, and energy extraction from the
rotating hole, as in the Blandford \& Znajek scenario, all appear to
play a role in different circumstances.  To the extent that numerical
simulations can provide a window on reality, it is emerging once again
that while simplified theoretical models shed light on aspects of
reality, ``all of the above'' seems to be the answer to the question of
which mechanism is responsible for launching jets from an accretion
flow.

\acknowledgements{
The work of JFH and JPD was supported by NSF grant AST-0070979 and NSF
ITR grant PHY02-05155. JHK and SH were partially supported by NSF
grants AST-0205806 and  AST-0313031 (JHK and SH). The simulations were
carried out on the Bluehorizon system of NPACI.  We thank Charles
Gammie and Jonathan McKinney for the discussion of their results. JHK
is also grateful to the Institute of Astronomy, Cambridge for its
hospitality while this work was completed, and to the Raymond and
Beverly Sackler Fund for support during his visit there.}


\clearpage

\begin{deluxetable}{lrrrrrrrrrrrrrrr} 
\tablecolumns{16} 
\tablewidth{0pc} 
\tablecaption{\label{Eject_fraction}
  Ejection from Outer Boundary}
\tablehead{\colhead{Model} & 
\colhead{$M_0$} & 
\colhead{${\Delta M_{\rm i}}$} & 
\colhead{${\Delta M_{\rm b}}$} &
\colhead{${\Delta M_{\rm u}}$} & 
\colhead{$\eta_M$} &
\colhead{$E_0$} &
\colhead{${\Delta E_{\rm i}}$} &
\colhead{${\Delta E_{\rm b}}$} &
\colhead{${\Delta E_{\rm u}}$} & 
\colhead{$\eta_E$} &
\colhead{$L_0$} & 
\colhead{${\Delta L_{\rm i}}$} &
\colhead{${\Delta L_{\rm b}}$} & 
\colhead{${\Delta L_{\rm u}}$} &
\colhead{$\eta_L$}}
\startdata 
KD0 & 156  & 24.9 & 1.14 & 0.17 & 0.7 
    & 154  & 22.7 & 1.14 & 0.22 & 1.0 
    & 902  & 77.1 & 6.85 & 0.44 & 0.6\\
KDI & 258  & 36.1 & 2.70 & 0.60 & 1.7 
    & 255  & 32.8 & 2.69 & 0.94 & 2.9
    & 1489 & 95.0 & 17.2 & 3.02 & 3.2 \\
KDP & 291  & 17.9 & 1.86 & 0.87 & 4.8 
    & 286  & 15.0 & 1.85 & 1.63 & 11
    & 1652 & 33.5 & 9.12 & 4.10 & 12\\
KDE & 392  & 14.4 & 8.35 & 4.78 & 33 
    & 386  & 9.34 & 8.33 & 9.67 & 104
    & 2255 & 14.1 & 55.2 & 23.0 & 163 \\
\\
\hline
\\
KDPi & 151  & 5.73 & 0.48 & 0.32 & 5.6 
     & 148  & 4.79 & 0.47 & 0.54 & 11
     & 835  & 10.8 & 2.34 & 1.52 & 14\\
KDPl & 291  & 10.2 & 0.88 & 0.87 & 8.5 
     & 286  & 8.33 & 0.88 & 1.52 & 18
     & 1652 & 19.2 & 4.21 & 3.68 & 19\\
SFR  & 741  & 285  & 8.76 & 1.36 & 0.5 
     & 726  & 272  & 8.73 & 1.67 & 0.6
     & 3484 & 1102 & 46.0 & 2.19 & 0.2\\
SF0  & 1374 & 469  & 24.6 & 6.78 & 1.4 
     & 1346 & 441  & 24.5 & 6.88 & 1.6
     & 6058 & 1455 & 113  & 24.7 & 1.7\\
SFP  & 2308 & 427  & 72.3 & 30.2 & 7.1
     & 2263 & 369  & 72.1 & 38.4 & 10
     & 9729 & 759  & 289  & 116  & 15\\
\enddata 
\end{deluxetable}

\clearpage

\begin{deluxetable}{lrrrr} 
\tablecolumns{5} 
\tablewidth{0pc} 
\tablecaption{\label{mass_table}
  Mass flux in the inner region}
\tablehead{\colhead{Model} & 
\colhead{${\Delta M_{\rm hole}}$} & 
\colhead{${\Delta M_{\rm jet}}$} & 
\colhead{${\Delta M_{\rm r = 15}}$} & 
\colhead{$M_{\rm final}$}}
\startdata 
KD0 & 0.161  & 0.00147 & 0.184 & 0.0194\\
KDI & 0.141  & 0.00275 & 0.171 & 0.0259\\
KDP & 0.062  & 0.00361 & 0.101 & 0.0281\\
KDE & 0.038  & 0.01400 & 0.115 & 0.0545\\
\enddata 
\end{deluxetable}
\clearpage

\clearpage

\clearpage

\clearpage

\clearpage

\clearpage

\clearpage

\clearpage

\end{document}